# Alan Turing's Legacy:
# Info-Computational Philosophy of Nature

Gordana Dodig-Crnkovic[1]

**Abstract.** Alan Turing's pioneering work on computability, and his ideas on morphological computing support Andrew Hodges' view of Turing as a natural philosopher. Turing's natural philosophy differs importantly from Galileo's view that the book of nature is written in the language of mathematics (The Assayer, 1623). Computing is more than a language of nature as computation produces real time physical behaviors. This article presents the framework of Natural Info-computationalism as a contemporary natural philosophy that builds on the legacy of Turing's computationalism. Info-computationalism is a synthesis of Informational Structural Realism (the view that nature is a web of informational structures) and Natural Computationalism (the view that nature physically computes its own time development). It presents a framework for the development of a unified approach to nature, with common interpretation of inanimate nature as well as living organisms and their social networks. Computing is understood as information processing that drives all the changes on different levels of organization of information and can be modeled as morphological computing on data sets pertinent to informational structures. The use of info-computational conceptualizations, models and tools makes possible for the first time in history the study of complex self-organizing adaptive systems, including basic characteristics and functions of living systems, intelligence, and cognition.

## 1 Turing and Natural Philosophy

Andrew Hodges [1] describes Turing as a Natural philosopher: "He thought and lived a generation ahead of his time, and yet the features of his thought that burst the boundaries of the 1940s are better described by the antique words: natural philosophy." Turing's natural philosophy differs from Galileo's view that the book of nature is written in the language of mathematics (The Assayer, 1623). Computation is not just a language of nature; it is the way nature behaves. Computing differs from mathematics in that computers not only calculate numbers, but more importantly they can produce real time physical behaviours.

Turing studied a variety of natural phenomena and proposed their computational modeling. He made a pioneering contribution in the elucidation of connections between computation and intelligence and his work on morphogenesis provides evidence for natural philosophers' approach. Turing's 1952 paper on morphogenesis [2] proposed a chemical model as the basis of the development of biological patterns such as the spots and stripes that appear on animal skin.

Turing did not originally claim that the physical system producing patterns actually performs computation through morphogenesis. Nevertheless, from the perspective of info-computationalism, [3,4] argues that morphogenesis is a process of morphological computing. Physical process, though not computational in the traditional sense, presents natural (unconventional), physical, morphological computation.

An essential element in this process is the interplay between the informational structure and the computational process – information self-structuring. The process of computation implements physical laws which act on informational structures. Through the process of computation, structures change their forms, [5]. All computation on some level of abstraction is morphological computation – a form-changing/form-generating process, [4].

In this article, info-computationalism is identified as a new philosophy of nature providing the basis for the unification of knowledge from currently disparate fields of natural sciences, philosophy, and computing. An on-going development in bioinformatics, computational biology, neuroscience, cognitive science and related fields shows that in practice biological systems are currently already studied as information processing and are modelled using computation-theoretical tools [6,7,8].

Denning declares: "Computing is a natural science" [9] and info-computationalism provides plenty of support for this claim. Contemporary biologists such as Kurakin [10] also add to this information-based naturalism, claiming that "living matter as a whole represents a multiscale structure-process of energy/matter flow/circulation, which obeys the empirical laws of nonequilibrium thermodynamics and which evolves as a self-similar structure (fractal) due to the pressures of competition and evolutionary selection". [11, p5]

## 2 Universe as Informational Structure

The universe is, from the metaphysical point of view, "nothing but processes in structural patterns all the way down" [12, p228]. Understanding patterns as information, one may infer that information is a fundamental ontological category. The ontology is scale-relative. What we know about the universe is what we get from sciences, as "special sciences track real patterns" [12, p242]. This idea of an informational universe coincides with Floridi's Informational Structural Realism [13,14]. We know as much of the world as we explore and cognitively process:

*"Reality in itself is not a source but a resource for knowledge. Structural objects (clusters of data as relational entities) work epistemologically like constraining affordances: they allow or invite certain constructs (they are affordances for the*

---

[1] School of Innovation, Design and Engineering, Mälardalen University, Sweden. Email: gordana.dodig-crnkovic@mdh.se

*information system that elaborates them) and resist or impede some others (they are constraints for the same system), depending on the interaction with, and the nature of, the information system that processes them."* [13, p370].

Wolfram [15] finds equivalence between the two descriptions – matter and information:

*"[M]atter is merely our way of representing to ourselves things that are in fact some pattern of information, but we can also say that matter is the primary thing and that information is our representation of that. It makes little difference, I don't think there's a big distinction – if one is right that there's an ultimate model for the representation of universe in terms of computation."* [16, p389].

More detailed discussion of different questions of the informational universe, natural info-computationalism including cognition, meaning and intelligent agency is given by Dodig Crnkovic and Hofkirchner in [17].

## 3 The Computing Universe – Naturalist Computationalism

Zuse was the first to suggest (in 1967) that the physical behavior of the entire universe is being computed on a basic level, possibly on cellular automata, by the universe itself, which he referred to as "Rechnender Raum" or Computing Space/Cosmos. Consequently, Zuse was the first pancomputationalist (natural computationalist), [18]. Chaitin in [19, p.13] claims that the universe can be considered to be a computer "constantly computing its future state from its current state, constantly computing its own time-evolution account!" He quotes Toffoli, pointing out that "actual computers like your PC just hitch a ride on this universal computation!"

Wolfram too advocates for a pancomputationalist view [15], a new dynamic kind of reductionism in which the complexity of behaviors and structures found in nature are derived (generated) from a few basic mechanisms. Natural phenomena are thus the products of computation processes. In a computational universe new and unpredictable phenomena emerge as a result of simple algorithms operating on simple computing elements such as cellular automata, and complexity originates from the bottom-up emergent processes. Cellular automata are equivalent to a universal Turing Machine. Wolfram's critics remark, however, that cellular automata do not evolve beyond a certain level of complexity; the mechanisms involved do not produce evolutionary development. Wolfram meets this criticism by pointing out that cellular automata are models and as such surprisingly successful ones. Also Fredkin [20] in his Digital philosophy builds on cellular automata, suggesting that particle physics can emerge from cellular automata. For Fredkin, humans are software running on a universal computer.

Wolfram and Fredkin, in the tradition of Zuse, assume that the universe is, on a fundamental level, a discrete system, and is thus suitably modelled as an all-encompassing digital computer. However, the computing universe hypothesis (natural computationalism) does not critically depend on the discreteness of the physical world, as there are digital as well as analog computers. On a quantum-mechanical level, the universe performs computation on characteristically dual wave-particle objects [21], i.e. both continuous and discrete computing. Maley [22] demonstrates that it is necessary to distinguish between analog and continuous, and between digital and discrete representations. Even though typical examples of analog representations use continuous media, this is not what makes them analog. Rather, it is the relationship that they maintain with what they represent. Similar holds for digital representations. The lack of proper distinctions in this respect is a source of much confusion on discrete vs. continuous computational models.

Moreover, even if in some representations it may be discrete (and thus conform to the Pythagorean ideal of number as a principle of the world), computation in the universe is performed at many different levels of organization, including quantum computing, bio-computing, spatial computing, etc. – some of them discrete, others continuous. So computing nature seems to have a use for both discrete and continuous computation, [23].

## 4 Information Processing Model of Computation

Computation is nowadays performed by computer systems connected in global networks of multitasking, interacting devices. The classical understanding of computation as syntactic mechanical symbol manipulation performed by an isolated computer is being replaced by the information processing view by Burgin, [24]. Info-computationalism adopts Burgin definition of computation as information processing.

In what follows, I will focus on explaining this new idea of computation, which is essentially different from the notion of context-free execution of a given procedure in a deterministic mechanical way. Abramsky summarizes this changing paradigm of computing as follows:

*"Traditionally, the dynamics of computing systems, their unfolding behaviour in space and time has been a mere means to the end of computing the function which specifies the algorithmic problem which the system is solving. In much of contemporary computing, the situation is reversed: the purpose of the computing system is to exhibit certain behaviour. (…)*

*We need a theory of the dynamics of informatic processes, of interaction, and information flow, as a basis for answering such fundamental questions as: What is computed? What is a process? What are the analogues to Turing completeness and universality when we are concerned with processes and their behaviours, rather than the functions which they compute?"* [25, p483]

According to Abramsky, there is a need for second generation models of computation, and in particular there is a need for process models such as Petri nets, Process Algebra, and similar. The first generation models of computation originated from problems of formalization of mathematics and logic, while processes or agents, interaction, and information flow are genuine products of the development of computers and Computer Science. In the second generation models of computation, previous isolated systems with limited interactions with the environment are replaced by processes or agents for which interactions with each other and with the environment are fundamental.

As a result of interactions among agents and with the environment, complex behaviour emerges. The basic building block of this interactive approach is the agent, and the fundamental operation is interaction. The ideal is the computational behaviour of an organism, not mechanical machinery. This approach works at both the macro-scale (such as processes in operating systems, software agents on the Internet, transactions, etc.) and on the micro-scale (from program implementation, down to hardware).

The above view of the relationship between information and computation presented in [25] agrees with ideas of info-computational naturalism of Dodig-Crnkovic [3] which are based on the same understanding of computation and its relation to information. Implementation of info-computationalism, interactive computing (such as, among others, agent-based) naturally suits the purpose of modelling a network of mutually communicating processes/agents, see [3,4,5].

## 5 Natural Computation

Natural computing is a new paradigm of computing which deals with computability in the natural world. It has brought a new understanding of computation and presents a promising new approach to the complex world of autonomous, intelligent, adaptive, and networked computing that has emerged successively in recent years. Significant for Natural computing is a bidirectional research [7]: as natural sciences are rapidly absorbing ideas of information processing, computing is concurrently assimilating ideas from natural sciences.

The classical mathematical theory of computation was devised long before global computer networks. Ideal, classical theoretical computers are mathematical objects and they are equivalent to algorithms, Turing machines, effective procedures, recursive functions or formal languages. Compared with new computing paradigms, Turing machines form the proper subset of the set of information processing devices, in much the same way as Newton's theory of gravitation presents a special case of Einstein's theory, or Euclidean geometry presents a limited case of non-Euclidean geometries, [5].

Natural/Unconventional computing as a study of computational systems includes computing techniques that take inspiration from nature, use computers to simulate natural phenomena or compute with natural materials (such as molecules, atoms or DNA). Natural computation is well suited for dealing with large, complex, and dynamic problems. It is an emerging interdisciplinary area closely related to artificial intelligence and cognitive science, vision and image processing, neuroscience, systems biology and bioinformatics, to mention but a few.

Computational paradigms studied by natural computing are abstracted from natural phenomena such as self-* attributes of living (organic) systems (including -replication, -repair, -definition and -assembly), the functioning of the brain, evolution, the immune systems, cell membranes, and morphogenesis.

Unlike in the Turing model, where the Halting problem is central, the main issue in Natural computing is the adequacy of the computational response (behaviour). The organic computing system adapts dynamically to the current conditions of its environments by self-organization, self-configuration, self-optimization, self-healing, self-protection and context-awareness. In many areas, we have to computationally model emergence which is not algorithmic according to Cooper [26] and Cooper and Sloman [27]. This makes the investigation of computational characteristics of non-algorithmic natural computation (sub-symbolic, analog) particularly interesting.

In sum, solutions are being sought in natural systems with evolutionary developed strategies for handling complexity in order to improve complex networks of massively parallel autonomous engineered computational systems. Research in theoretical foundations of Natural computing is needed to improve understanding of the fundamental level of computation as information processing which underlies all computing.

## 6 Information as a Fabric of Reality

*"Information is the difference that makes a difference. "* [29]

More specifically, Bateson's difference is the difference in the world that makes the difference for an agent. Here the world also includes agents themselves. As an example, take the visual field of a microscope/telescope: A difference that makes a difference for an agent who can see (visible) light appears when she/he/it detects an object in the visual field. What is observed presents a difference that makes the difference for that agent. For another agent who may see only ultra-violet radiation, the visible part of the spectrum might not bring any difference at all. So the difference that makes a difference for an agent depends on what the agent is able to detect or perceive. Nowadays, with the help of scientific instruments, we see much more than ever before, which is yet further enhanced by visualization techniques that can graphically represent any kind of data.

A system of differences that make a difference (information structures that build information architecture), observed and memorized, represents the fabric of reality for an agent. Informational Structural Realism [13] [30] argues exactly that: information is the fabric of reality. Reality consists of informational structures organized on different levels of abstraction/resolution. A similar view is defended in [12]. Dodig Crnkovic [3] identifies this fabric of reality (Kantian 'Ding an sich') as *potential information* and makes the distinction between it and *actual information* for an agent. Potential information for an agent is all that exists as not yet actualized for an agent, and it becomes information through interactions with an agent for whom it makes a difference.

Informational structures of the world constantly change on all levels of organization, so the knowledge of structures is only half the story. The other half is the knowledge of processes – information dynamics.

It is important to note the difference between the potential information (world in itself) and actual information (world for an agent). Meaningful information, which is what in everyday speech is meant by information, is the result of interaction between an agent and the world. Meaning is use, and for an agent information has meaning when it has certain use. Menant [31] proposes to analyze relations between information, meaning and representation through an evolutionary approach.

## 7 Info-Computationalism as Natural Philosophy

Info-computationalist naturalism identifies computational process with the dynamic interaction of informational structures. It includes digital and analog, continuous and discrete, as phenomena existing in the physical world on different levels of organization. Our present-day digital computing is a subset of a more general Natural computing. In this framework, computational processes are understood as natural computation, since information processing (computation) is not only found in human communication and computational machinery but also in the entirety of nature.

Information represents the world (reality as an informational web) for a cognizing agent, while information dynamics (information processing, computation) implements physical laws through which all the changes of informational structures unfold.

Computation, as it appears in the natural world, is more general than the human process of calculation modelled by the Turing machine. Natural computing takes place through the interactions of concurrent asynchronous computational processes, which are the most general representation of information dynamics [5].

## 8 Conclusions

Alan Turing's work on computing machinery, which provided the basis for artificial intelligence and the study of its relationship to natural intelligence, together with his computational models of morphogenesis, can be seen as a pioneering contribution to the field of Natural Computing and the Computational Philosophy of Nature. Today's info-computationalism builds on the tradition of Turing's computational Natural Philosophy. It is a kind of epistemological naturalism based on the synthesis of two fundamental cosmological ideas: the universe as informational structure (informationalism) and the universe as a network of computational processes (pancomputationalism/naturalist computationalism).

Information and computation in this framework are two complementary concepts representing structure and process, being and becoming. Info-computational conceptualizations, models and tools enable the study of nature and its complex, dynamic structures, and uncover unprecedented new possibilities in the understanding of the connections between earlier unrelated phenomena of non-living and living nature [28].